# Design and Implementation of a Fleet Management System Using Novel GPS/GLONASS Tracker and Web-Based Software

Hamed Saghaei

Department of Electrical Engineering, Faculty of Engineering
Shahrekord Branch, Islamic Azad University, Shahrekord - Iran.
Email: h.saghaei@iaushk.ac.ir, Tel Number: +983833361000-399

*Abstract*— Knowing where the vehicles are, what the drivers doing and monitoring every event in real time is the key parameters for a well-managed decision-making process. In this paper, a novel approach for control and monitoring of a fleet management system using three elements including GPS/GLONASS-based automatic vehicle locators (called Rad100), GPRS/SMS GSM cellular network and web-based software (called PayaRadyab) is proposed to show exact position of the desired vehicle on different maps and take detailed reports of the mission, travelled path, fuel consumption rate, speed limits, and other necessary information according to the customers' requests. The most significant features of the proposed system are its global covering, high accuracy of positioning, easy operation by the user at any location, and easy energy management. In this study, I have designed and fabricated more than 50 Rad100 trackers and also programmed a web-based PayaRadyab software in which their performance and accuracy have been confirmed by the practical results in different conditions.

*Keywords*— *Automatic Vehicle Locator (AVL), Fleet Management System (FMS), Global Positioning System (GPS), Global Navigation Satellite System (GLONASS), General Packet Radio Service (GPRS).*

## I. INTRODUCTION

Fleet management system is used for monitoring different kinds of motor vehicles such as cars, vans, trucks, aircraft (planes, helicopters etc.), ships as well as rail cars [1]. It has numerous applications such as vehicle maintenance [2], vehicle tracking and diagnostics [3], improving driver performance [4], speed control and fuel management [5]. Fleet Management is a function which allows companies which rely on transportation in business to remove or minimize the risks associated with vehicle investment, improving efficiency, productivity and reducing their overall transportation and staff costs, providing 100% compliance with government legislation (duty of care) [6] and much more. These functions can be dealt with by either an in-house fleet management department or an outsourced fleet management provider [1]. The number of fleet management units deployed in commercial fleets in Europe will grow from 1.5 million units in 2009 to 4 million in 2014 [7]. Even though the overall penetration level is just a few percent, some segments such as road transport will attain adoption rates above 31 percent [8]. All major truck manufacturers on the European market offer OEM telematics solutions as a part of their product portfolio. Mercedes-Benz [9], Volvo and Scania launched their first products in the 1990s and followed by MAN in 2000, Renault Trucks in 2004, DAF Trucks in 2006 and IVECO in 2008 [10]. The products are all supporting the FMS standard and can generally be deployed in mixed fleets even if some functionality can be brand-specific [11]. A major trend in 2008 and 2009 has been the announcement of solutions for remote downloading of digital tachograph data and more advanced functionalities for eco-driving [12].

Our intelligent system for supervision, control, and management of vehicle fleet, called PayaRadyab, has been developed, programmed, and implemented recently. In this system, we employed the latest programmable integrated circuits (ICs) technology as well as the most powerful state of the art software. Using this system, the managers or users can conduct online web-based tracking of vehicle fleet on server based maps. Moreover, this system makes available detailed reports of the mission, location, fuel consumption rates, speed limits, and other required information according to customers' requests. The most significant features of this system are its global covering, high positioning accuracy, easy operation by the user at any location, and easy energy management. The proposed system comprises different hardware and software parts which are presented in the following sections with more details. In Section II, Advantages of fleet management systems are discussed and also Fleet management system is presented as detailed information. PayaRadyab software is presented in Section III and finally, the paper is closed with a conclusion in Section IV.

## II. ADVANTAGES OF FLEET MANAGEMENT SYSTEMS

Nowadays, management and planning as well as control and monitoring of various activities in transportation sectors are considered to be extremely important at a global level and can lead to further developments in economic, social, and even political fields. The fleet management system was developed based on the above approach and made available to applicants and various users so that they could plan the mission, service, and operational vehicle based on actual and exact objective information for the purpose of providing better control and monitoring with due attention to the content of the assigned missions. The following advantages can be enumerated for this system

- Management of vehicles fuel consumption based on daily, monthly, and annual reports provided for a particular vehicle or a group of vehicles.
- Efficient and exact management of the vehicle fleet and increase supervision capabilities

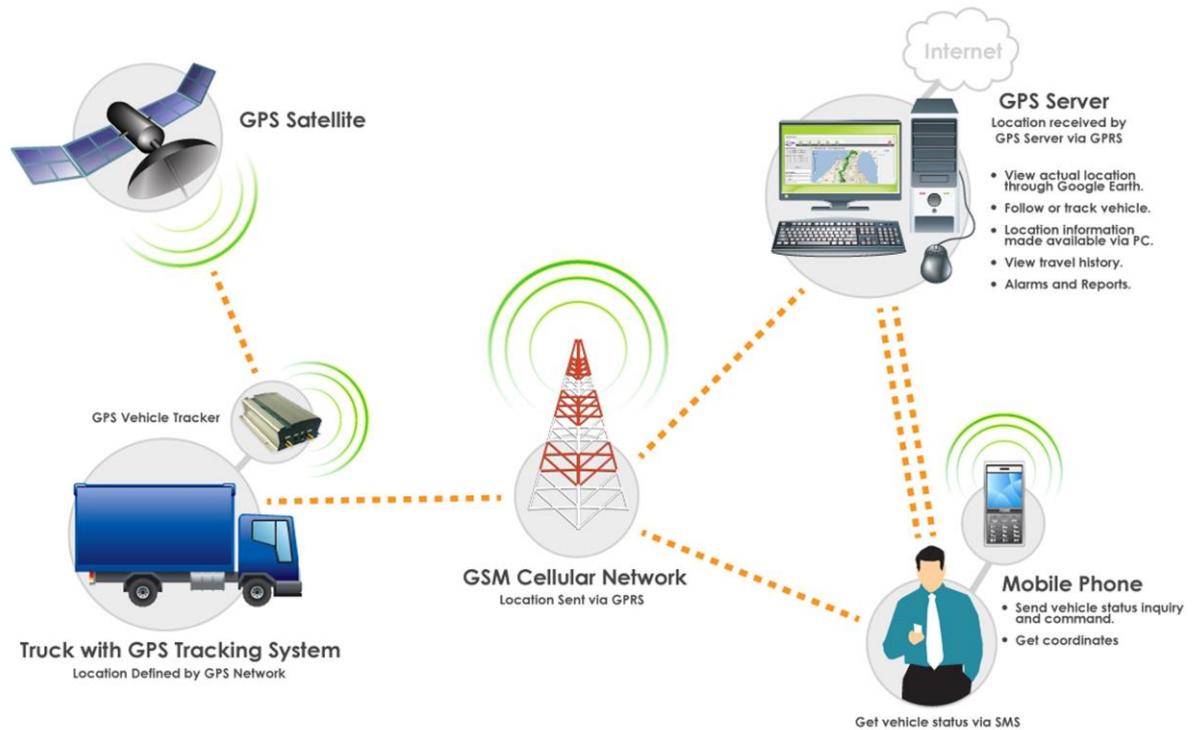

Fig. 1. Fleet management system based on AVL, GPS satellites, and GSM cellular network.

- Promotion of system efficiency and considerable reduction of control and monitoring costs compared with traditional operator-based supervisory systems
- Receiving exact performance and operation information from the vehicles
- Increase the fleet management system efficiency
- Considerable reduction of driving violations during in-service periods
- Increase the customer satisfaction and staff transparency
- Possibility of evaluating the performance of the affiliated organizations
- Standardization of the implemented concepts and forms within the executive organizations for the purpose of correcting the existing methods and preventing subjective trends to govern various processes.

*A. Fleet management system*

PayaRadyab system shown in Fig. 1 comprises a device called tracker and a software. Our designed and fabricated tracker, Rad100, is used for receiving satellite waves and transmitting the received information to the desired web server using a GPRS connection. When Rad100 receives the satellite waves at least from 4 GPS/Glonass satellites, by exact calculations and necessary operations, the position of the device can be achieved and stored on its memory card. It will send such information to a web server as either synchronous (online) using the GPRS or SMS platforms in the GSM communication network, or asynchronous (for later offline loading into the program). By receiving the data from Rad100, our designed software, PayaRadyab, stores the vehicle positions in its SQL database and subsequently displays them on various online maps. PayaRadyab is provided as a web-based version, allowing the users to receive the necessary information directly from Rad100 everywhere they like just by an internet connection and a common web browser such as IE, Mozilla or google chrome.

*B. Intelligent GPS/GLONASS tracker*

Rad100 shown in Fig. 2 is a terminal with GPS/GLONASS and GSM connectivity that is able to determine the vehicle's coordinates and transfer them via GPRS or SMS of the GSM network toward a defined web server. It is a very reliable hardware to minimize the theft risk provided that external GPS antenna is not accessible of the thief. It not only calculates the exact position of the vehicle but also supports some digital inputs/outputs that can be used for another installed security system on the vehicle. The GPS information picked up by Rad100 can be dealt with in one of two ways. For active trackers, it is instantly relayed through the GSM network to PayaRadyab software which provides a live picture of where the vehicle is located. With passive trackers, the journey information is recorded on the memory card of Rad100 and can be downloaded at a later date, usually when the vehicle returns to base. Rad100 can show the direction and speed of a vehicle as well as its location. Some features of Rad100 are organized as follows:

**GPS features**
- 32 channel GPS/GLONASS receiver with -161 dBm sensitivity
- NMEA, GGA, GGL, GSA, GSV, RMC, VTG protocol compatible

**GPRS features**
- GSM frequency of 850MHz, 900MHz, 1800MHz, and 1900MHz
- GPRS class 10 and SMS (text, data)

**Special features**
- CPU ARM CORTEX-M3
- 16 MB internal Flash memory (120 days data storing) with the SD card support
- Built-in movement sensor
- Power supply: from 6 to 32V
- Protection against overcurrent, short circuiting and earth faults
- Integrated scenarios:
- ECO driving (ratings of acceleration, braking, cornering based on accelerometer)
- Over-speeding, Authorized driving (50 iButton keys), Immobilizer
- USB port, and 4 status LEDs
- 4 digital and 2 analog inputs with a dedicated digital input for Ignition and 4 open collector outputs
- Any element event triggers (external sensor, input, speed, temperature, etc.)
- Highly configurable data acquisition and sending
- Real Time tracking and Smart data acquisition (based on time, distance, angle, ignition and events
- Sending acquired data via GPRS (TCP/IP and UDP/IP protocols)
- Smart algorithm of GPRS connections (GPRS traffic saving)
- Operating in roaming networks (preferred GSM providers list)
- Events on I/O detection and sending via GPRS or SMS.
- 150 geofence zones (rectangular, circle or triangle)
- Different sleep modes (idle, normal sleep, deep sleep)
- Authorized number list for remote access
- Firmware update over GPRS or USB port
- Configuration its settings over GPRS, SMS or USB port
- TCP/IP or UDP/IP protocol support,
- IP, domain and Port settings.
- Panic alert, speed alert and towing alert
- Highly configurable functionality
- Hermetic enclose IP67

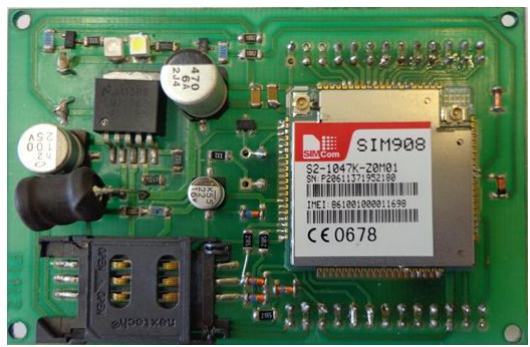

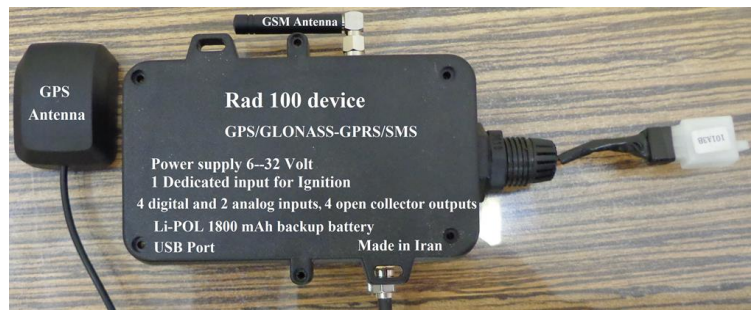

Fig. 2. Rad100 Hardware and box with embedded Rad100 tracker

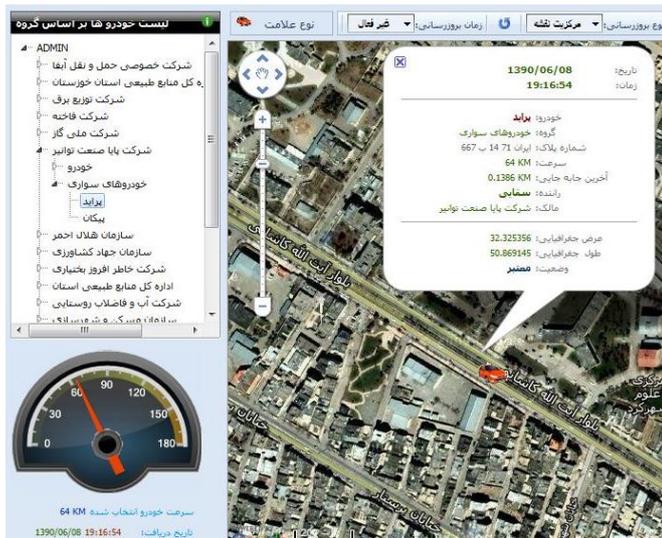

Fig. 3: The latest vehicle position on the map using the proposed native software (PayaRadyab Software in the Persian language).

- Location (periodic/on-demand)
- GSM/GPRS jamming detection
- Driver ID (iButton, Keypad) and Emergency key
- Simple and optimized text mechanism for individual operation
- Li-Pol rechargeable battery 3.7V, 1800 mAh
- External GSM antenna and external GPS/GLONASS antenna
- One-wire, two-wire interface protocol, SPI, RS232 and CAN interfaces
- Weight (including battery): 100 g, Storage temperature -40 to 85° C
- Operating temperature -20 to 70° C, Max. Relative humidity 90±5%
- Power consumption @ 12Vdc:
- Max: 70-100 mA (avg) Peak: 400 mA
- Low power mode (GPS off): < 10 mA
- Sleep mode (GPS off, modem off): < 3 mA

In the case of no coverage for GSM and GPS systems (which occurs in most vehicles), the proposed system can store the required information for up to 4 months and automatically transmit the stored data from the hardware to the server as soon as signal coverage is restored.

### III. PAYARADYAB SOFTWARE AND ITS APPLICATIONS

PayaRadyab, a native software in the Persian language, is designed for online monitoring and control of Rad100, installed on the vehicle. It is programmed to fulfill three parts of applications including:

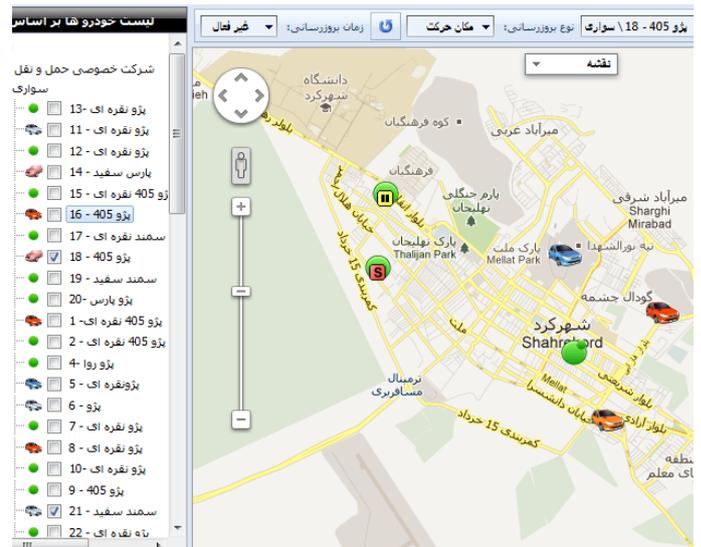

Fig. 4: The last vehicles' positions on the map using PayaRadyab Software.

- *Short-term applications*: In this case, the effective outputs from the system are available to the user(s) in a short period of time for quick decision making.
- *Supervisory and management applications*: In this application, long-term statistical reports are extracted for long-term management and planning.
- *Automotive consumables control*: According to the received data from different sensors and mileage of the tracker installed on the car, PayaRadyab software gives comprehensive information about the status of automotive consumables. Thus, it considerably reduces peripheral costs and, simultaneously, prevents probable vehicle damage and depreciation.

#### A. Short-term applications

Short-term applications of PayaRadyab software are organized as follows:

- Display the latest position and online monitoring of the vehicle(s)
- Speed limit control and sending related warnings
- Find the mileage and vehicle heading
- Monitor the stop and start points beside their durations
- Find the nearest vehicle to our desired position

Figs. 3 and 4 show the latest position of a vehicle and a group of vehicles on different kinds of online maps, respectively, using PayaRadyab software. Users can access to their necessary information just by click on the icon of the desired vehicle after they were logged in to the system.

Fig. 5 illustrates the nearest vehicle(s) to the desired position of the user which is applied for an emergency. In other words, through displaying the last position of a certain vehicle, the system can identify the nearest vehicle to the desired location and assign the relevant mission to that vehicle, thus providing targeted vehicle use towards accomplishing the mission and reducing the fuel consumption at the same time. Another application is crisis control and management. In the event of various crises such as fire, the proposed system can be applied to display the positions of a group of vehicles and subsequently dispatch the nearest vehicles or motorcycles to that area. Thus, the information provided by PayaRadyab software can be used for crisis management. The traveled vehicle paths and the direction of movements are important parts of every fleet management system. As shown in Fig. 6, the path and the direction of the desired vehicle can be displayed by PayaRadyab software in terms of different features such as speed limit and time interval.

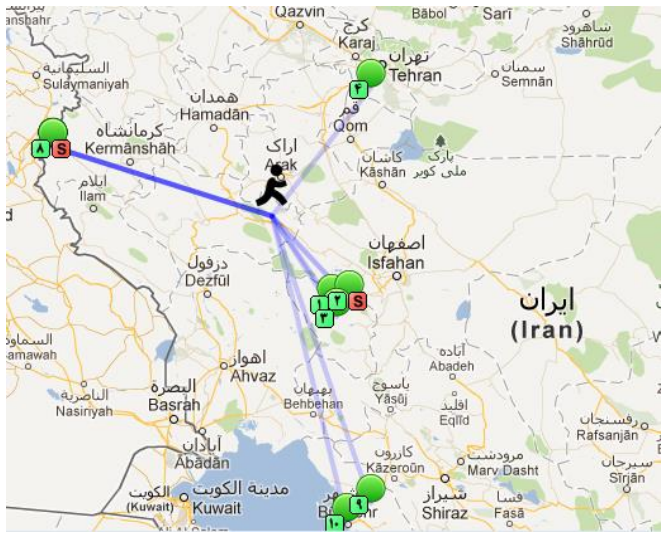

Fig. 5. The distance of every selected vehicle to our desired location.

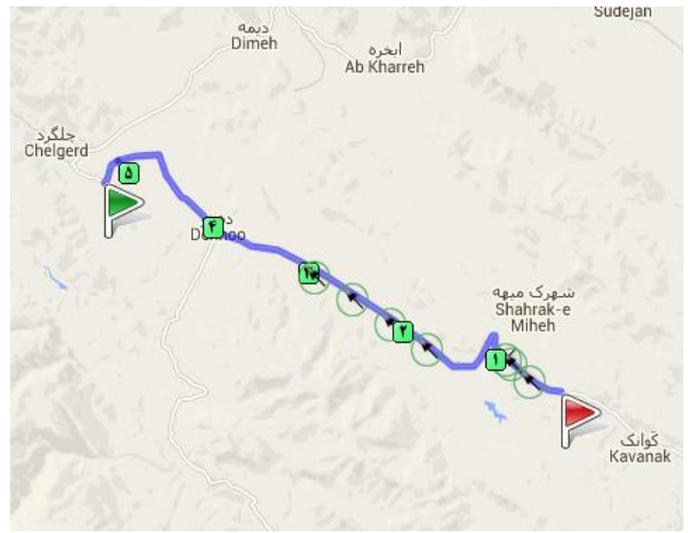

Fig. 6. The travelled path and the direction of the desired vehicle during the time interval

### B. Supervisory and management applications

Using PayaRadyab software, the user can extract various comprehensive and comparative reports for better supervisory and management such as mileage, mission, average and maximum speed, and fuel consumption of either a vehicle or a group of vehicles in a defined interval of time.

Comprehensive reports can be used for making decisions about budget allocation and performance evaluation based on the mileage and fuel consumption calculated for the vehicles while comparative ones are used for comparison of mentioned reports between two vehicles daily and monthly.

One of the reported problems by the managers of different organizations is the payment of vehicle-related expenses. To solve it, by filling out the "mission form" provided by the PayaRadyab system, the respective private company can specify the vehicle use during missions. These forms can be extracted from the system every month or according to the required date. The mileage calculation of by each vehicle can then be performed by PayaRadyab software and the due payments forwarded to the relevant organization. In this part of the software, for long-distance missions, the fuel consumption for the different paths leading to the same destination can be calculated and the path with the minimum consumption selected for the future mission. PayaRadyab software also presents different kinds of reports about fuel consumption versus the velocity of the vehicle. For example, if the light vehicle travels 100 km smooth path at speeds less than 70 km/hr., fuel consumption would increase and finally it increases our expenses. It also reports another report about remaining prepaid SIM card charge which is a remarkable feature provided by the software. The average and maximum speed of the vehicle can be given by the software too. Figs. 7, 8 and 9 show different extracted reports of PayaRadyab software.

In Fig. 7(a) the traveled path versus date is depicted as a bar graph. It can be observed that on the first day, the vehicle traveled around 577 km. However, it is 414 km for $30^{th}$. The maximum traveled path by that vehicle is around 1071 km in $17^{th}$. Fig. 7(b) shows fuel consumption of the vehicle in each day of the $11^{th}$ month too in which Rad100 extract its value from the car and send it back to our server. As you see in this figure, there are relatively a direct relation between traveled path and fuel consumption.

In Figs. 8(a) and 8(b) both traveled path and fuel consumption are depicted as bar graphs monthly from $6^{th}$ month to $12^{th}$ month. As you see in these figures, the most traveled path of the vehicle has been done in the $9^{th}$ month.

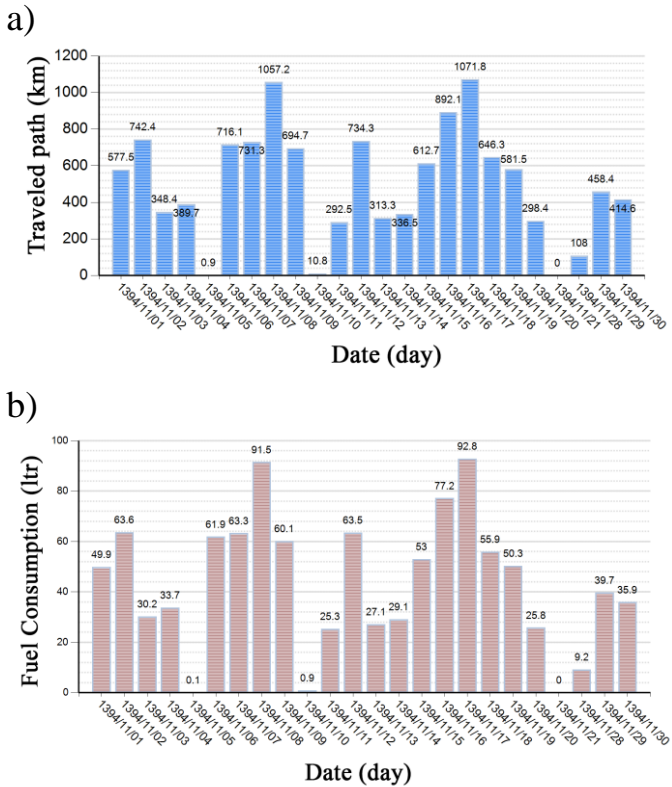

Fig. 7. a) Traveled path and b) Fuel consumption of a vehicle in a month based on daily report part of the PayaRadyab software.

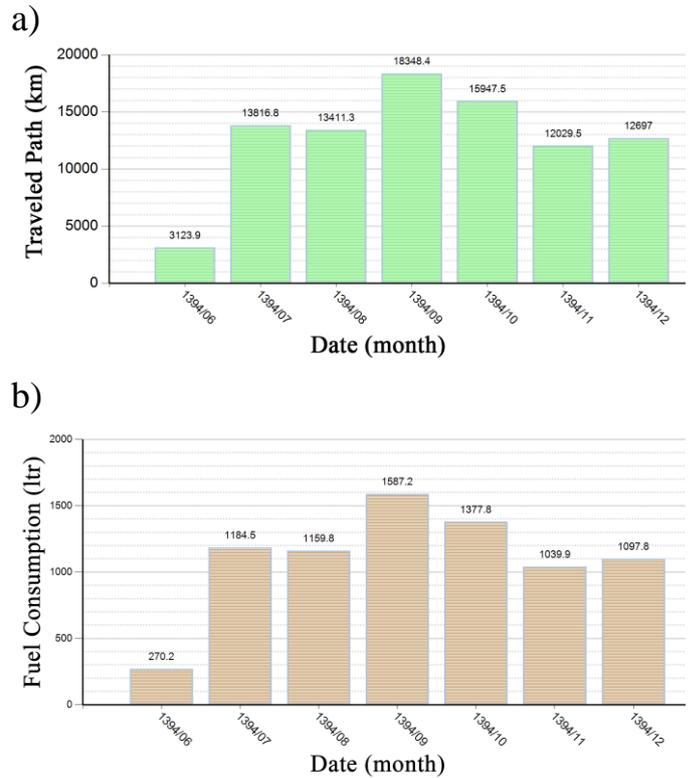

Fig. 8. a) Traveled path and b) Fuel consumption of a vehicle in half a year versus each month based on monthly report part of the PayaRadyab software.

The maximum fuel is proportionally consumed in that month.

Figs. 9(a) and 9(b) show a comparative study of traveled path and fuel consumption of that vehicle in two desired months in which the blue and orang bars are related to 9[th] and 10[th] months. As you observe in this figure, in first day of the 9[th] month the traveled path is less than that day in the 10[th] month. Using these reports and others necessary information of PayaRadyab software, the organization managers can monitor and manage their vehicles in different missions.

*C. Automotive consumables control*

A vehicle has many consumable items (e.g. engine oil, brake fluid, spark plug, belt, air and oil filters, etc.) which must be replaced at regular intervals or vehicle performance can suffer greatly, leading to excessive fuel consumption and even complete breakdown. To avoid such incidents, the consumables must be controlled based on the mileage and replaced appropriate times. PayaRadyab software alarms inform users about them.

PayaRadyab Software facilitates your replacement and procurement process by providing you with the needed data to make informed decisions regarding the condition of your equipment. You can set up your replacement schedules at the individual equipment level or at the class level for a group of equipment. Once setup, PayaRadyab software will provide budgetarily and replacement reports to assist in the replacement process. These reports can also be exported to an excel spreadsheet in order to make the procurement process easier. PayaRadyab Software contains very comprehensive history reporting for each piece of fleet equipment. Want to know when the last brake job was done? It's in there! Want to know how many of these alternators have been installed and on which vehicles? It's in there! Whether you are looking for a basic historical or cost report or a more complex report - PayaRadyab gives you many reports to pull the data out.

Without a doubt, fleet security is the most important area of all. Driver and public safety are of course paramount, and anyone running a fleet of vehicles has to make this their prime concern. As an additional security feature, some fleet management services offer customers access to vehicular control mechanisms. Security systems located within the vehicle can enable employers to gradually and safely reduce the speed of, and ultimately stop, a vehicle within their fleet. This security feature is particularly useful for companies operating fleets of large vehicles which have the potential to cause significant damage in the event that they are stolen or become rogue (where the vehicle travels without an occupant or the occupant is unable to stop the vehicle from within). To get the maximum benefits, it is essential that both the fleet manager and the drivers have a good working knowledge of the tracking software.

PayaRadyab software can run on a number of different hardware platforms and databases. It is supported on Windows XP, Vista, Windows 7, or Windows 8 as the client and Windows Server 2003, Windows Server 2008, or Windows 2012 as the file and database server. It runs on MS/SQL Server 2005, 2008 or 2012. I studied power control process for 3$^{rd}$ and 4$^{th}$ generations of wireless cellular communication systems for the future work [13-16]. Also, optical communication devices used for telecommunication data transmission are studied in [17-20].

IV. CONCLUSION

We studied and presented a novel approach for control and monitoring of GPS-based fleet management system. This proposed system consists of a tracker, Rad100, which is installed on the vehicle and web-based software that should be installed at the server side. The GPS information picked up by Rad100 will be sent to PayaRadyab software by the GPRS connection which provides a live picture of where the vehicle is located and also detailed reports of the mission, traveled path, fuel consumption rate, speed limits, and other necessary information can be extracted from PayaRadyab Software. With this software, in addition to equipment replacement, the user can also factor in inflationary affects as well as any up-fitting that might be required when the equipment is replaced.